\documentclass{article}

\usepackage{arxiv}

\usepackage{algorithmic}
\usepackage{algorithm}
\usepackage[utf8]{inputenc} 
\usepackage[T1]{fontenc}    
\usepackage{hyperref}       
\usepackage{url}            
\usepackage{booktabs}       
\usepackage{amsmath,amsfonts}       
\usepackage{nicefrac}       
\usepackage{microtype}      
\usepackage{lipsum}
\usepackage{graphicx}
\graphicspath{ {./images/} }

\title{Autonomous Circular Drift Control for 4WD-4WS Vehicles Without Precomputed Drifting Equilibrium}

\author{
 Yue Xiao \\
  Intelligent Transportation System Research Center\\
  Wuhan University of Technology\\
  Wuhan, China 430063 \\
  \texttt{xiao\_yue@whut.edu.cn} \\
   \And
 Yi He \\
  Intelligent Transportation System Research Center\\
  Wuhan University of Technology\\
  Wuhan, China 430063 \\
  \texttt{heyi@whut.edu.cn} \\
  \And
 Yaqing Zhang \\
  Intelligent Transportation System Research Center\\
  Wuhan University of Technology\\
  Wuhan, China 430063 \\
  \texttt{zhangyaqing@whut.edu.cn} \\
\And
 Xin Lin \\
  Intelligent Transportation System Research Center\\
  Wuhan University of Technology\\
  Wuhan, China 430063 \\
  \texttt{298402@whut.edu.cn} \\
  \And
 Ming Zhang \\
  Hubei Sanjiang Aerospace Wanshan Special Vehicle Co., Ltd.\\
  Xiaogan, China 430063 \\
  \texttt{78939016@qq.com} \\
}

\begin{document}
\maketitle
\begin{abstract}
Under extreme conditions, autonomous drifting enables vehicles to follow predefined paths at large slip angles, significantly enhancing the control system's capability to handle hazardous scenarios. Four-wheel-drive and four-wheel-steering (4WD-4WS) vehicles, which have been extensively studied, offer superior path-following precision and enhanced maneuverability under challenging driving conditions. In this paper, a hierarchical drifting controller is proposed for 4WD-4WS vehicles to track both path and velocity without relying on precomputed drifting equilibrium. The controller is structured into two layers: a trajectory tracking layer and an actuator regulation layer. The first layer generates the desired tire forces in the vehicle body frame, while the second layer converts these desired tire forces into steering angle commands and torque commands for the front and rear motors. The effectiveness and robustness of the proposed controller are validated through simulation. 
\end{abstract}


\section{Introduction}
There are various types of cornering for autonomous vehicles, with most approaches placing greater emphasis on keeping the vehicle within the stable region, which typically corresponds to a low sideslip angle. However, operating beyond the stability boundaries does not necessarily render the vehicle uncontrollable. Under extreme driving conditions, the controller should maximize the utilization of friction forces between the ground and tires to induce vehicle drift, enabling the vehicle to quickly adjust its body posture.

\begin{figure}[!t]
\centering
\includegraphics[width=3.3in]{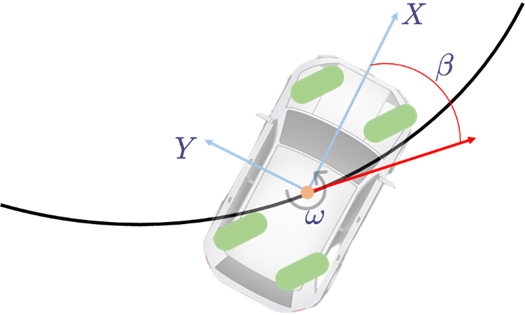}
\caption{Vehicle drifting state.}
\label{fig_1}
\end{figure}

When the vehicle is drifting, the sideslip angle is typically large and opposite to the direction of the yaw rate. Simultaneously, the wheels steer in the opposite direction to stabilize the sideslip angle, as shown in Fig. \ref{fig_1}.

Extensive research has been conducted by numerous scholars on drifting control for different types of vehicles. Early studies, such as those presented in \cite{hoffman2008using} and\cite{bobier2013staying}, primarily focused on analyzing drifting equilibrium using a two-degree-of-freedom (2DoF) vehicle model. By analyzing the phase plane, the existence of unstable equilibrium was proven. Velenis et al. \cite{velenis2011steady} used a three-degree-of-freedom (3DoF) model and the Pacejka tire model to obtain the numerical solution for the drifting equilibrium of rear-wheel-drive (RWD) and front-wheel-steering (FWS) vehicles. In contrast, Velenis \cite{velenis2011fwd} also focused on front-wheel-drive (FWD) vehicles, calculating equilibrium using a vehicle model with driven front wheels, while the rear wheels are 'locked' at zero angular rate through the application of the handbrake. Remarkably, the vehicle can also be controlled to drift while reversing, and the reverse driving car is modeled as a FWD vehicle with rear-wheel-steering (RWS) control \cite{khan2018steady}. Overall, most drifting con-trollers are designed according to the analysis of the drifting equilibrium of RWD vehicles, based on a simplified bicycle model \cite{baur2019experimentally}, \cite{bardos2020implementation}. Even so, the drift equilibrium solution also depends on the accuracy of the tire model and the efficiency of nonlinear equation-solving algorithms.

Recent years have seen growing academic interest in advanced vehicle control, especially for 4WD and 4WS systems. Compared to traditional vehicles, they improve low-speed maneuverability and high-speed stability, offering superior control and driving adaptability \cite{ruiz2019front}, \cite{xu2019improving}. Yu et al. \cite{yu2023vehicle} exploit the characteristics of vehicle input couplings to make the controller comprehensibly track the desired sideslip only when the sideslip dynamic is unstable, thereby fully unlocking the potential of 4WD electric vehicles. A hierarchical control method is employed to separately regulate curvature and sideslip, enabling the 4WD vehicle to drift on low-friction roads \cite{yang2022hierarchical}.

Although rear-wheel steering is often associated with improved maneuverability and oversteer, studies suggest that at higher velocities, implementing RWS to induce a subtle understeer effect can paradoxically enhance vehicle stability by mitigating unstable movements \cite{brabec2016stability}. This implies that 4WS, initially for improving understeer, can be repurposed for con-trolled instability in drift scenarios. Actively manipulating 4WS enables precise drift initiation and maintenance, thereby unlocking advanced levels of vehicle control in extreme maneuvering \cite{zhu2024nmpc}.

Compared to traditional vehicles, a major challenge in the drift control of 4WD and 4WS vehicles is that the excessive number of control inputs makes it difficult to analyze drift equilibrium. The steady-state equations based on the 3DoF model can determine at most the solutions for three variables. Typically, the wheel steering angle and velocity are set as constant, and it is assumed that the rear tire is saturated \cite{tian2023multi}. Milani et al. \cite{milani2022vehicle} argue that FWD vehicles cannot achieve steady-state drifting unless the rear tire slips are actively manipulated. In contrast, RWD and 4WD vehicles are found to exhibit unstable drifting equilibrium. 

To tackle the aforementioned challenge, some researchers have explored autonomous drift control methods that avoid relying on drifting equilibrium. In \cite{joa2019drift}, the desired yaw rate is designed to be operated in two states to resolve the oscillation problem. The novelty of this work lies in the path tracking algorithm without knowledge of drift equilibrium. How-ever, the paper lacks a clear explanation of the control algorithm, hinders reproducibility, and exhibits significant steady-state error. \cite{cai2020high}, \cite{yin2020self}, \cite{zhao2022vehicle}, \cite{chen2023dynamic} demonstrate that the reinforcement learning method appears to be an effective approach for handling complex control tasks, as it does not require accurate dynamic models. 

Another major challenge is the nonlinear tire model. Directly incorporating it into the vehicle model increases system complexity. Consequently, many studies employ hierarchical controllers \cite{yu2023vehicle}, \cite{chen2023dynamic}, especially for 4WD and 4WS vehicles. Nevertheless, in the lower layer, tire forces from the upper layer are converted into wheel steering angles, potentially introducing errors due to model mismatches and tire nonlinearities.

Moving on to control methods are used in the drifting control, apart from the above-mentioned reinforcement learning methods, linear quadratic regulator (LQR) \cite{velenis2011steady}, \cite{khan2018steady}, \cite{bardos2020implementation}, model predictive control (MPC) \cite{bellegarda2021dynamic}, \cite{lee2022real}, \cite{zhou2022learning}, sliding mode control (SMC) \cite{hou2022autonomous}, robust control \cite{xu2023narrow}, \cite{xu2021robust}, etc. have been widely explored. Indeed, a diverse range of control strategies can be employed to stabilize the vehicle around its drifting equilibria. Among them, MPC is often preferred for its ad-vantages in handing various constraints. 

Drift control can be categorized into two main types: transient drift and sustained drift \cite{chen2023dynamic}. Transient drift aims to achieve specific objectives, such as obstacle avoidance, by controlling the vehicle through a series of drifting states. In contrast, sustained drift focuses on maintaining the vehicle in an unstable equilibrium for an extended period, with an em-phasis on drift dynamics and theoretical analysis.

In this article, a drifting control algorithm is proposed for 4WD-4WS vehicles to track trajectory. The main contributions are as follows:

\begin{enumerate}
  \item A hierarchical controller is designed to address the challenge of unattainable drifting equilibrium, comprising two layers.
  \item To mitigate errors introduced by the inverse tire model, a novel MPC algorithm that accounts for input disturbances is adopted. This approach reduces the steady-state error to an acceptable range.
  \item Simulation results from MATLAB/Simulink and CarSim validate the effectiveness of the proposed controller in tracking a circular trajectory and maintaining drift equilibrium states, while also exploring the differences between 4WD-4WS vehicles and oth-er types of vehicles.
\end{enumerate}

This article is structured as follows: Section II introduces a 3DoF bicycle model for 4WD-4WS vehicles and the Magic Tire Formula. Section III details the controller design, including the upper-layer MPC and lower-layer inverse tire model. Section IV presents simulations validating the control meth-od. Finally, the study is summarized, and future research directions are proposed.

\begin{table}[!t]
\caption{Description of Symbols\label{tab:symbol}}
\centering
\begin{tabular}{|c||c|}
\hline
Meaning & Symbol\\
\hline
Vehicle mass & $m$\\
Moment inertia about vertical axis & $I_z$\\
Radius of wheel & $r$ \\
Distance from CG to axle & $a$, $b$\\
Wheel steering angle & $\delta_f$, $\delta_r$ \\ 
Sideslip angle & $\beta$ \\
Yaw rate & $\omega$ \\
Vehicle velocity & $v$ \\
Yaw angle & $\varphi$ \\
Road adhesion coefficient & $\mu$ \\
Tire slip angle & $\alpha_f$, $\alpha_r$ \\
\hline
\end{tabular}
\end{table}

\section{Task description and data construction}

A 4WD-4WS vehicle is equipped with independent drive motors on all four wheels and steering capability on both front and rear wheels. The symbols used throughout the paper are described in Table \ref{tab:symbol}.
\begin{figure}[!t]
\centering
\includegraphics[width=3.3in]{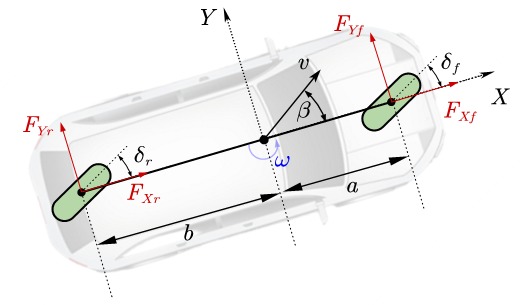}
\caption{3DoF bicycle model of 4WD-4WS vehicle.}
\label{fig_2}
\end{figure}

\subsection{Vehicle Dynamics Model}
What distinguishes this article from previous research is the tire forces in the 3DoF bicycle model are decomposed on the x 
and y-axis of the vehicle body, as shown in Fig. \ref{fig_2}, with the aim of simplifying the model. Thus, the model’s inputs are the longitudinal force of the front axle $F_{Xf}$, the longitudinal force of the rear axle$F_{Xr}$, the lateral force of the front axle$F_{Yf}$, and the lateral force of the rear axle $F_{Yr}$. 

The governing equations for the model are as follows:
\begin{equation}
\label{dynamics model}
\left\{
\begin{aligned}
\dot{\beta} &= \frac{\left( F_{Yf}+F_{Yr} \right) \cos \beta -\left( F_{Xf}+F_{Xr} \right) \sin \beta}{mv}-\omega \\	
\dot{\omega} &=\frac{1}{I_z}\left( aF_{Yf}-bF_{Yr} \right)\\	\dot{v}&=\frac{\left( F_{Xf}+F_{Xr} \right) \cos \beta +\left( F_{Yf}+F_{Yr} \right) \sin \beta}{m}\\
\end{aligned}
\right.
\end{equation}

\subsection{Tire Model}
\begin{figure}[!t]
\centering
\includegraphics[width=3.3in]{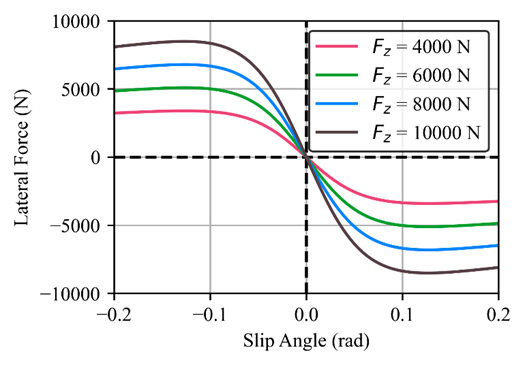}
\caption{Lateral tire force curve.}
\label{fig_3}
\end{figure}

The Magic Tire Formula\cite{pacejka1992magic} provides a more accurate and com prehensive representation of tire behavior under large slip angles, effectively capturing the nonlinear relationship between slip angle and lateral force, making it particularly suitable for extreme driving conditions like drifting. This article adopts a simplified Magic Tire Formula to model the lateral force of the steering tire: 
\begin{equation}
    \label{lateral tire model}
    F_{yf} = \mu F_{zf} \sin(C \arctan (B \alpha_f))
\end{equation}
\noindent where $F_{yf}$is the lateral force of the front axle, $F_{zf}$ is the vertical load of the front axle, $B$ and $C$ are constant parameters representing the stiffness factor and shape factor, respectively. Fig. \ref{fig_3} shows the relationship between the lateral force and tire slip angle.

The slip angle of the front tire and the rear tire are determined as in \eqref{tire slip angle} and Fig. \ref{tire slip angle}.
\begin{equation}
\label{tire slip angle}
    \begin{aligned}
        \alpha_f = \arctan(\frac{v\sin\beta+a\omega}{v\cos\beta})-\delta_f \\
        \alpha_r = \arctan(\frac{v\sin\beta-b\omega}{v\cos\beta})-\delta_r
    \end{aligned}
\end{equation}

\begin{figure}[!t]
\centering
\includegraphics[width=3.3in]{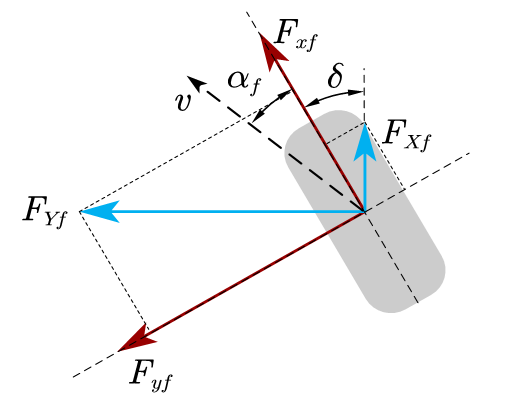}
\caption{Tire force and slip angle.}
\label{fig_4}
\end{figure}

The saturated tire force-slip relationship is commonly used to model rear tires in the design of automated drifting controllers. In contrast, this article avoids the assumption to achieve greater generality.

\subsection{Trajectory Tracking Model}
\begin{figure}[!t]
\centering
\includegraphics[width=3.3in]{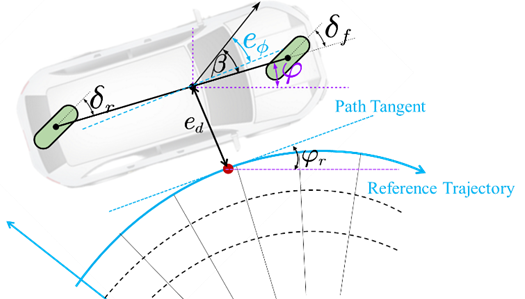}
\caption{Vehicle trajectory tracking model.}
\label{fig_5}
\end{figure}

Four error states of the vehicle relative to the reference trajectory are identified, which are the lateral error $e_d$, the tangential error $e_\phi$, the velocity error $e_v$, and the yaw rate error $e_\omega$. As shown in Fig. \ref{fig_5}, the tangential error $e_\phi$ is defined as \eqref{tan error}. 

\begin{equation}
\label{tan error}
    e_\phi = \varphi-\varphi_r + \beta
\end{equation}

The model is expressed as the 
\begin{equation}
\label{tracking model}
\left\{
\begin{aligned}
\dot{e_d} &= v \sin e_\phi \\	
\dot{e_\phi} &= \omega - \kappa \frac{v \cos e_\phi}{1-e_d\kappa} +\beta\\	
\dot{e_v}&=\dot{v} \\
\dot{e_\omega} &= \dot{\omega}
\end{aligned}
\right.
\end{equation}

\noindent With $(F_{Xf}, F_{Xr}, F_{Yf}, F_{Yr})^T$ as the control variables $u$, by incorporating \eqref{dynamics model}, the complete state-space equations can be easily derived and linearized:
\begin{equation}
\label{discreate state space}
    x_{k+1} = Ax_k + Bu_k
\end{equation}

\noindent where $A$, $B$ are the state matrix and input matrix, respectively.
\begin{equation}
    \label{Ak}
    A=\left[ \begin{matrix}	
    0&		v_r&		0&		0\\	
    {\kappa _r}^2v_r&		0&		-\kappa _r&		0\\	
    0&		0&		0&		0\\	
    0&		0&		0&		0\\
    \end{matrix} \right]
\end{equation}

\begin{equation}
    \label{Bk}
    B = 
    \renewcommand{\arraystretch}{2.2}  
    \begin{bmatrix}
        0 & 0 & 0 & 0 \\
        -\dfrac{\sin \beta}{mv} & -\dfrac{\sin \beta}{mv} & \dfrac{\cos \beta}{mv} & \dfrac{\cos \beta}{mv} \\
        \dfrac{\cos \beta}{m} & \dfrac{\cos \beta}{m} & \dfrac{\sin \beta}{m} & \dfrac{\sin \beta}{m} \\
        0 & 0 & \dfrac{a}{I_z} & -\dfrac{b}{I_z}
    \end{bmatrix}
\end{equation}

\noindent where $v_r$ is the reference velocity, $\kappa_r$ is the curvature of the reference point. 

\begin{figure}[!t]
\centering
\includegraphics[width=3.3in]{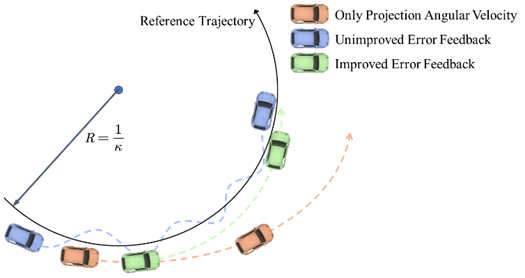}
\caption{Comparison of different desired yaw rates.}
\label{Comparison of different desired yaw rates}
\end{figure}

Determining the desired yaw rate becomes challenging in the absence of a precomputed drifting equilibrium. Moreover, relying solely on the angular velocity of the vehicle's projection point on the curve is insufficient, as it decreases as the vehicle deviates from the reference trajectory, contradicting common sense. To address this, the desired yaw rate pro-posed in \cite{joa2019drift} has been improved as shown in \eqref{fd yaw rate}, where a simple error feedback mechanism is employed to accelerate the vehicle into a drifting state. Fig. \ref{Comparison of different desired yaw rates} shows the differences among the three designs.
\begin{equation}
    \label{fd yaw rate}
    \omega _r=-\kappa _r\frac{v_r\cos e_{\phi}}{1-e_d\kappa _r}-k_1e_d-k_2e_{\phi}
\end{equation}

\section{DRIFT CONTROLLER DESIGN}

\begin{figure}            
  \centering                     
\includegraphics[width=0.95\textwidth]{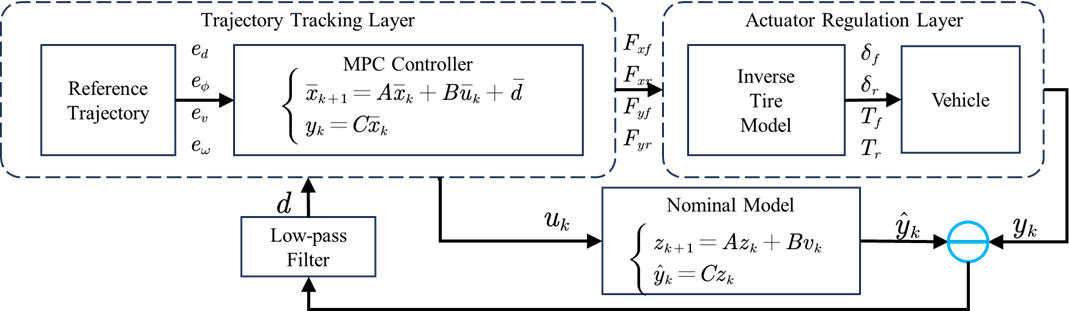} 
  \caption{The overall architecture of the hierarchical drifting controller.} 
  \label{system}                
\end{figure}

Section III begins by introducing the controller design, which is structured into two layers: the upper layer employing an MPC controller and the lower layer utilizing an in-verse tire model. The overall architecture of the controller is illustrated in Fig. \ref{system}.

In this section, the fundamentals of the MPC controller are first introduced, highlighting its role in trajectory tracking. Then, an inverse tire model based on the Newton-Raphson method is proposed to compute the wheel steering angles. Finally, an enhanced version of the MPC controller, incorporating input disturbance considerations, is discussed to improve performance.

\subsection{Trajectory Tracking layer}
Based on the discrete state-space equations in \eqref{discreate state space}, an incremental form of the MPC controller can be established:
\begin{equation}
    \label{MPC model}
    \left\{
    \begin{aligned}
    &\bar{x}_{k+1}=\bar{A}x_k+\bar{B}\Delta u_k \\	
    &y_k=C\bar{x}_k
    \end{aligned}
    \right.
\end{equation}

The predictive state variables are calculated sequentially using the set of future control parameters:
\begin{equation}
    \label{predictive function}
    Y_k=\varPsi \bar{x}_{k|k}+\varTheta \Delta U_k
\end{equation}

\noindent with
\begin{equation*}
    Y_k = \begin{bmatrix} 
        y_{k+1|k} \\ 
        y_{k+2|k} \\ 
        \vdots \\ 
        y_{k+Np|k}
    \end{bmatrix}, 
    \Delta U_k = \begin{bmatrix} 
        \Delta u_{k|k} \\ 
        \Delta u_{k+1|k} \\ 
        \vdots \\ 
        \Delta u_{k+Nc-1|k}
    \end{bmatrix}, 
    \varPsi = \begin{bmatrix} 
        C\bar{A} \\ 
        C\bar{A}^2 \\ 
        \vdots \\ 
        C\bar{A}^{Np}
    \end{bmatrix}, 
\end{equation*}
\begin{equation*}
    \varTheta =\left[ \begin{matrix}	C\bar{B}&		0&		\cdots&		0\\	C\bar{A}\bar{B}&		C\bar{B}&		\cdots&		0\\	\vdots&		\vdots&		\ddots&		\vdots\\	C\bar{A}^{Np-1}\bar{B}&		C\bar{A}^{Np-2}\bar{B}&		\cdots&		C\bar{A}^{Np-Nc}\bar{B}\\\end{matrix} \right]. 
\end{equation*}

\noindent where $Np$ and $Nc$ are the prediction horizon and control horizon, respectively.

The cost function is as follows, which includes both the error cost and the control increment cost, and can be transformed into a standard quadratic programming (QP) problem:
\begin{equation}
    \label{optimal question}
    J=\sum_{i=1}^{Np}{\left\| Y_k-Y_{ref} \right\| _{Q}^{2}}+\sum_{i=0}^{Nc-1}{\left\| \Delta U_k \right\| _{R}^{2}}
\end{equation}

\noindent where $Q$ and $R$ are weighting matrices.

\begin{figure}[!t]
\centering
\includegraphics[width=3.3in]{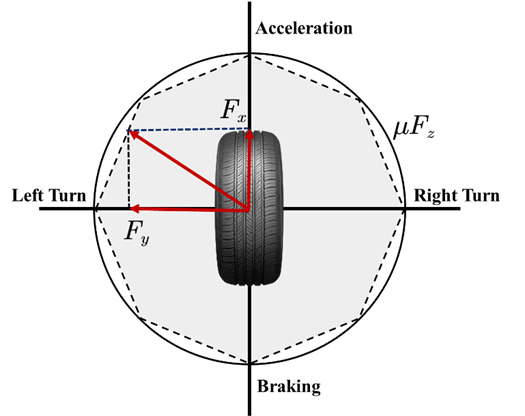}
\caption{Tire friction circle.}
\label{Tire friction circle}
\end{figure}

The total force generated by the wheels is constrained by the friction circle, which is defined by the friction coefficient and the vertical load. However, this constraint is inherently nonlinear, presenting challenges for efficient computation. To address this, it is reformulated as a set of linear constraints using an octagonal approximation method shown in Fig. \ref{Tire friction circle}, as proposed in this article.

To find the optimal $\Delta U_k$, the cost function $J$ is minimized, formulated as a QP problem that includes inequality constraints and variable bounds:
\begin{equation}
    \label{eq:qp_problem}
    \begin{aligned}
        & \min_{\Delta U_k} \quad J(\Delta U_k) = \frac{1}{2} \Delta U_k^\top \mathbf{H} \Delta U_k + \mathbf{f}^\top \Delta U_k \\
        & \mathrm{subject~to} \quad 
        \begin{cases}
            \mathbf{A}_{\mathrm{ieq}} \Delta U_k \leq \mathbf{b}_{\mathrm{ieq}} \\
            \Delta U_{\min} \leq \Delta U_k \leq \Delta U_{\max}
        \end{cases}
    \end{aligned}
\end{equation}

It is worth noting that the sideslip angle   is considered constant during each control period, which implies that the accuracy of the tracking model will decrease when the reference trajectory is not sufficiently smooth.

The parameters used for the design of the MPC controller are listed in Table \ref{tab:para}. The weight matrices and gains were fine-tuned through a trial-and-error approach.

\begin{table}[!t]
\caption{VEHICLE AND CONTROL PARAMETERS\label{tab:para}}
\centering
\begin{tabular}{|c||c|}
\hline
Symbol & Value\\
\hline
$m$                & $1600~\mathrm{kg}$\\
$I_z$              & $1536.7~\mathrm{kg \cdot m^2}$\\
$a$                & $1.015~\mathrm{m}$ \\
$b$                & $1.895~\mathrm{m}$\\
$T$                & $0.05~\mathrm{s}$\\ 
$B$                & $-11.52$ \\
$C$                & $1.62$ \\
$r$                & $0.325$ \\
$Np$               & $30$ \\
$Nc$               & $8$ \\
$\delta_{max}$     & $35^\circ$ \\
$\Delta F_{Xmax}$  & $1500~\mathrm{N/s}$ \\
$\Delta F_{Ymax}$  & $14000~\mathrm{N/s}$ \\
$k_1$              & $0.15$ \\
$k_2$              & $0.1$ \\
$\gamma$           & $0.98$ \\
$Q$                & $\mathrm{diag[2900,2000,1000,7500]}$\\
$R$                & $\mathrm{diag[1,1,0.01,0.01]}$\\
\hline
\end{tabular}
\end{table}

\subsection{Actuator Regulation Layer}
In this layer, the steering angles and motor torques are calculated to generate the tire forces specified by the upper layer. The longitudinal inverse tire model is not discussed here for simplicity, as the effect of tire slip ratio is neglected in this article.

Focusing on the front axle in the following sections, the expression for the tire slip angle as a function of lateral force is derived from \eqref{lateral tire model} and presented as \eqref{inverse tire model}.
\begin{equation}
    \label{inverse tire model}
    \alpha _f=\frac{1}{B}\tan \left( \frac{1}{C}\mathrm{arc}\sin \left( \frac{F_{yf}}{\mu F_{zf}} \right) \right) 
\end{equation}

Due to the presence of the   term in the equation, the slip angle $\alpha_f$ is constrained to the linear region, as   is only defined for values within the range $[-1, 1]$, limiting the scope of $\alpha_f$.

Fig. \ref{inverse tire model} illustrates how multiple slip angle solutions are obtained from the inverse Magic Formula tire model. However, since the slip angle is bounded within a certain range, multiple solutions only arise when the lateral force $F_{yf}$ is very close to its maximum value $\mu F_{zf}$. Therefore, the complete expression is given by \eqref{eq:multiple_solutions}.

\begin{equation}
    \label{eq:multiple_solutions}
    \begin{aligned}
        f(F_{yf}) &= \left\{ \alpha_f, \alpha_f', \alpha_f'' \right\} \\
        &= \begin{cases} 
            \frac{1}{B} \tan\left( \frac{1}{C} g(F_{yf}) \right), & \text{if } |F_{yf} - \mu F_{zf}| < \epsilon \\
            \frac{1}{B} \tan\left( \frac{1}{C} (\pi - g(F_{yf})) \right), & \text{if } |F_{yf} + \mu F_{zf}| < \epsilon
        \end{cases}
    \end{aligned}
\end{equation}
where
\begin{equation*}
    g(F_{yf}) = \arcsin\left( \frac{F_{yf}}{\mu F_{zf}} \right).
\end{equation*}

\begin{figure}[!t]
\centering
\includegraphics[width=3.3in]{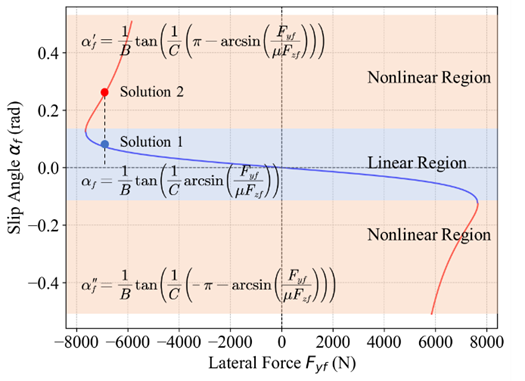}
\caption{Multiple solutions of the inverse tire model.}
\label{Multiple solutions of the inverse tire model}
\end{figure}

\noindent where $\epsilon$ is a pre-defined threshold. A reasonable choice of   can influence the computational efficiency of the wheel steering angle calculation, as discussed below. 

The relationship between the lateral and longitudinal tire forces and the vehicle-body frame forces from the upper layer is as follows:
\begin{equation}
    \label{coord trans}
    \begin{aligned}
         F_{xf}&=F_{Xf}\cos \delta _f+F_{Yf}\sin \delta _f\\
         F_{yf}&=-F_{Xf}\sin \delta _f+F_{Yf}\cos \delta _f
    \end{aligned}
\end{equation}

 By combining \eqref{tire slip angle}, a nonlinear objective function for the iterative solution of the wheel steering angle is obtained:
\begin{equation}
    \label{L_function}
    \begin{aligned}
        L(\delta_f) &= f\bigl( -F_{\mathrm{Xf}} \sin \delta_f + F_{\mathrm{Yf}} \cos \delta_f \bigr) \\
        &\quad - \arctan\left( \frac{v \sin \beta + a \omega}{v \cos \beta} \right) + \delta_f
    \end{aligned}
\end{equation}

The Newton-Raphson method is a fast and efficient iterative method used to solve nonlinear equations or systems numerically. Compared to other methods like the bisection method, secant method, and fixed-point iteration, it offers significant advantages in convergence speed and computational efficiency. It’s essential to consider the possibility of tire forces exceeding the friction circle constraint during iteration. The iterative framework is shown in Algorithm \ref{alg:newton_raphson}.
\begin{algorithm}[!t]
\caption{Front Steering Angle Iteration Based on Newton-Raphson Method}
\label{alg:newton_raphson}
\begin{algorithmic}[1]
\REQUIRE $\mu, F_{zf}, \delta_{\max}, \text{tolerance}, \text{max\_iter}$
\ENSURE $\delta_{f,\text{new}}$
\STATE Initialize $i \leftarrow 0,\ \delta_f \leftarrow 0$
\WHILE{$i < \text{max\_iter}$}
    \STATE Compute $L(\delta_f)$ and $L'(\delta_f)$
    \IF{$F_{yf} > \mu F_{zf}$}
        \STATE $F_{yf} \leftarrow \mu F_{zf}$  \COMMENT{Friction limit constraint}
    \ENDIF
    \STATE Calculate step: $\Delta \delta = L(\delta_f)/L'(\delta_f)$
    \STATE $\delta_{f,\text{new}} \leftarrow \delta_f - \Delta \delta$
    \STATE Apply saturation: 
    $\delta_{f,\text{new}} \leftarrow \max(-\delta_{\max}, \min(\delta_{f,\text{new}}, \delta_{\max}))$
    \IF{$|\delta_{f,\text{new}} - \delta_f| < \text{tolerance}$}
        \RETURN $\delta_{f,\text{new}}$  \COMMENT{Convergence achieved}
    \ENDIF
    \STATE Update: $\delta_f \leftarrow \delta_{f,\text{new}}$
    \STATE $i \leftarrow i + 1$
\ENDWHILE
\RETURN $\delta_{f,\text{new}}$  \COMMENT{Return best estimate after max iterations}
\end{algorithmic}
\end{algorithm}

After determining the steering angle, the longitudinal tire force can be calculated accordingly. A simple approach is adopted to compute the front torque command $T_f$. 
\begin{equation}
    T_f = (F_{Xf}\cos \delta_f + F_{Yf} \sin \delta_f)r
\end{equation}

\subsection{Error Compensation}
\begin{figure}[!t]
\centering
\includegraphics[width=3.3in]{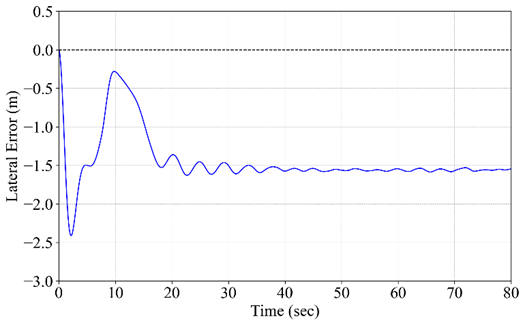}
\caption{Simulation result of drifting controller without error compensation.}
\label{Simulation result of drifting controller without error compen-sation}
\end{figure}

Fig. \ref{Simulation result of drifting controller without error compen-sation} shows the tracking performance of a 30-meter circular trajectory at a planned speed of 10 m/s without error compensation, revealing a steady-state error of approximately 1.5 meters. 

\begin{figure}[!t]
\centering
\includegraphics[width=3.3in]{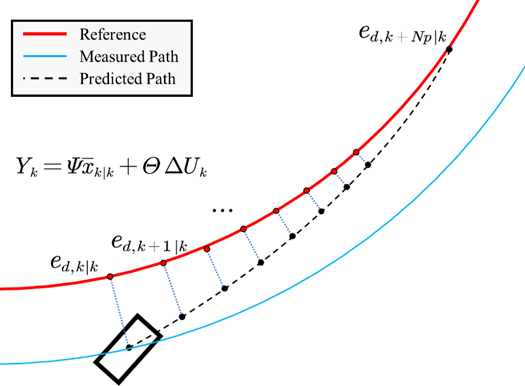}
\caption{Vehicle trajectory tracking under MPC control at 40 seconds: comparison of reference path, predicted path, and measured path.}
\label{Vehicle trajectory tracking}
\end{figure}

Regarding the cause of steady-state error in MPC, it is generally believed to be due to an overly simplified model \cite{chu2022trajectory}. Fig. \ref{Vehicle trajectory tracking} shows the predicted lateral errors at 40 seconds derived from the prediction equations \eqref{predictive function}. It can be observed that the lateral error decreases over time, indicating some level of error in the model itself.

Specifically, in the inverse tire model discussed earlier, the sideslip angle and the yaw rate are taken from the last timestep, which leads to inaccurate tire slip angles as obtained in \eqref{tire slip angle}. This means that the tire forces obtained by implementing the wheel steering angle in the lower layer differ from the optimal forces computed by MPC in the upper layer.

\begin{figure}[!t]
\centering
\includegraphics[width=3.3in]{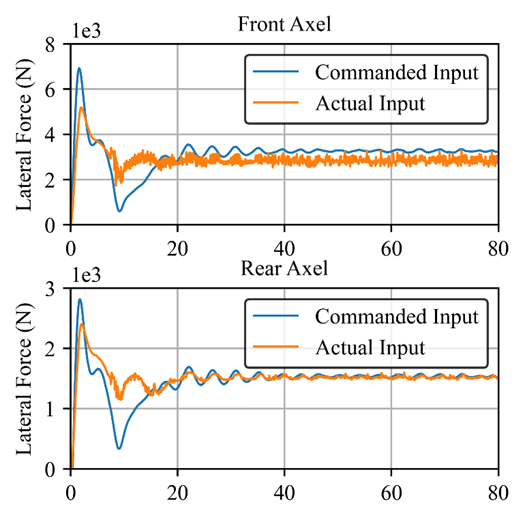}
\caption{Commanded and actual lateral forces on front and rear axles.}
\label{command and actual}
\end{figure}

Fig. \ref{command and actual} shows the commanded and actual lateral forces on the front and rear axles. Small errors are observed on the front axle, with lateral force deviations reaching several a few hundred Newtons. This discrepancy is the primary cause of the steady-state error in the control system. Resolving this issue might be possible by obtaining real-time control deviations and applying feedforward compensation. However, it is unfortunate that most vehicles are not equipped with tire force sensors.

Models with input disturbances, as shown in \eqref{state space with disturbance}, typically have control methods based on disturbance observer design \cite{yu2018mpc}, or use a PID controller to compensate for steady-state error \cite{chu2022trajectory}, both of which introduce additional complexity. 

\begin{equation}
    \label{state space with disturbance}
    \begin{aligned}
        x_{k+1} &= Ax_k+B(u_k+u_{bias}) \\
        &= Ax_k + Bu_k +d_k
    \end{aligned}
\end{equation}

\noindent where $u_{bias}$ is the input disturbance.

Equation \eqref{predictive function} can be rewritten as:
\begin{equation}
    \label{predictive function with disturbance}
    Y_k=\varPsi \bar{x}_{k|k}+\varTheta \Delta U_k
\end{equation}

Incorporating $D$ directly into the cost function may lead to oscillations in the vehicle's behavior, as the disturbance $d_k$ is typically unknown or only estimated in real time. Including it directly in the cost function assumes perfect knowledge of $d_k$, which is unrealistic in most practical scenarios. This can result in inaccurate predictions of the system's future behavior.

\begin{figure}[!t]
\centering
\includegraphics[width=3.3in]{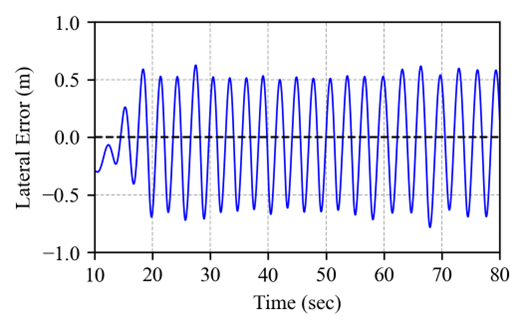}
\caption{Simulation results with   directly integrated into the cost function.}
\label{oscillation}
\end{figure}

Furthermore, MPC is designed to optimize control inputs within constraints, assuming the system evolves predictably. If $D$ is directly included, any inaccuracies in its estimation may cause the controller to generate overly aggressive or conservative control actions. This can lead to instability or degraded performance, as the controller may overreact to disturbances. As shown in Fig. \ref{oscillation}, the control exhibits severe oscillation issues.

Instead of directly incorporating $D$ into the cost function, a more robust approach is to treat $d_k$ as a bounded disturbance and design the MPC with robustness in mind. Techniques such as robust MPC or tube-based MPC can be used to account for the effects of $d_k$ while maintaining stability and constraint satisfaction.

However, tube-based MPC faces challenges due to its conservativeness, reliance on accurate disturbance bounds, and poor scalability in high-dimensional systems. Overly cautious tube design can reduce efficiency, while inaccurate disturbance bounds risk infeasibility or inefficiency. Additionally, increasing system dimensionality exponentially complicates tube design, making it impractical for large-scale applications with limited computational resources.

In this article, a method is devised to reduce the steady-state error while minimizing computational cost, which calculates the disturbance $d_k$ directly from the difference be-tween the true plant dynamics and the nominal model. Then, the disturbance is passed through a low-pass filter to eliminate high-frequency noise and is subsequently fed into the MPC module.

\begin{figure}[!t]
\centering
\includegraphics[width=3.3in]{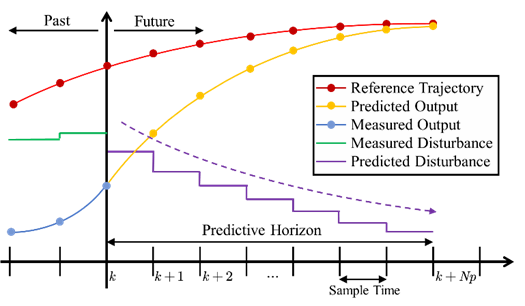}
\caption{System response under MPC control, illustrating the tracking performance, and disturbance rejection over time.}
\label{mpc}
\end{figure}

The key issue is that the disturbance is highly correlated with the control input $d_k$ and the vehicle’s state. Therefore, the confidence in   decreases over time. An attenuation co-efficient $\gamma$ is introduced to model this reduction, which is then incorporated into the MPC as presented in Fig. \eqref{mpc}. 

The vector $\mathbf{f}$ in \eqref{eq:qp_problem} is modified as shown in \eqref{qp's f}. 
\begin{equation}
\label{qp's f}
\mathbf{f}=2\varTheta ^T\bar{Q}\left( \varPsi \bar{x}_{k|k}+D_{corr} \right)
\end{equation}

\noindent where 
\begin{equation*}
    D_{corr}=\left[ \begin{array}{c}	\gamma d_k\\	\gamma^{2}\left( \bar{A}+I \right) d_k\\	\vdots\\	\gamma^{Np}\left( \sum_{i=0}^{Np}{\bar{A}^i} \right) d_k\\\end{array} \right]
\end{equation*}

It should be emphasized that the value of the coefficient $\gamma$ should be adjusted based on the smoothness of the reference trajectory and the sample time. For trajectories with sharp changes or longer sample times, smaller values are recommended.

In summary, the steady-state error stems from the tire force capture limitations. Therefore, achieving a solution demands balancing error reduction, robustness, and efficiency.

\section{MODEL AND SIMULATION VALIDATION}
Before conducting the simulation validation of the designed hierarchical controller, it is necessary to validate the established vehicle dynamics model to ensure the validity and reliability of the simulation results. Therefore, this section first introduces the process and results of model validation. Upon confirmation of the model's validity, simulation validation was conducted on MATLAB/Simulink and CarSim platforms. The simulation results demonstrate that the designed controller achieves satisfactory tracking of circular reference trajectories, including a 30 m radius circle, and a variable curvature circle with continuously increasing curvature. The vehicle starts from the origin with an initial velocity of 10 m/s.

\begin{figure}[!t]
\centering
\includegraphics[width=3.3in]{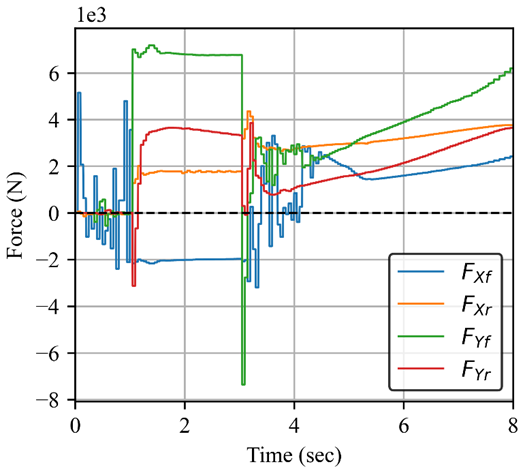}
\caption{Tire forces in the vehicle body coordinate system.}
\label{tire forces in the vehicle body}
\end{figure}

\subsection{Model Validation}
The vehicle dynamics model, \eqref{dynamics model}, effectively manifests inherent nonlinearities, thus necessitating high accuracy for robust representation. Drift simulation was conducted, employing double-step inputs on steering and driving torque. These inputs rapidly elicit system excitation, facilitating transient response observation and effective dynamic performance assessment. Specifically, at 1 second, the front and rear wheels were steered oppositely via a step input; at 2.5 seconds, a counter-steering step input was applied to the front wheels, rapidly inducing drift.

Fig. \ref{tire forces in the vehicle body} depicts the longitudinal and lateral forces on the front and rear axles, defined in the vehicle body coordinate system.

\begin{figure}[!t]
\centering
\includegraphics[width=3.3in]{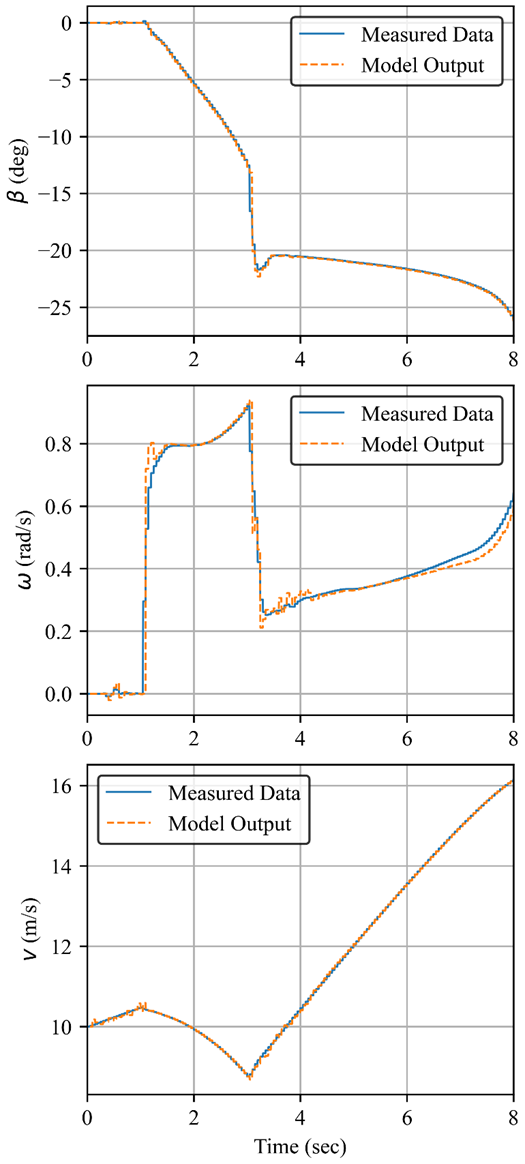}
\caption{Comparison of model predictions and CarSim simulations.}
\label{model val}
\end{figure}

Fig. \ref{model val} and Table \ref{tab:model error} jointly validate the effectiveness of the vehicle dynamics model. Fig. \ref{model val} Comparison of model predictions and CarSim simulations visually demonstrates the high consistency between model predictions and CarSim simulations through time-domain curve comparisons. This shows the model accurately replicates CarSim's vehicle dynamics characteristics. Table \ref{tab:model error} quantitatively presents the Root Mean Square Error (RMSE) and maximum error for each state variable, with all values being very low, further confirming the model's accuracy.

\begin{table}[!t]
\caption{MODEL ERROR ANALYSIS\label{tab:model error}}
\centering
\begin{tabular}{|c||c||c|}
\hline
State Variable & RMSE & Max Error\\
\hline
$\mathrm{Sideslip Angle\ (deg)}$ & 0.29 & 3.39\\
$\mathrm{Yaw Rate\ (rad/s)}$     & 0.04 & 0.3\\
$\mathrm{Vehicle Velocity\ (m/s)}$ & 0.04 & 0.15 \\
\hline
\end{tabular}
\end{table}

\subsection{Constant Curvature Circular}
\begin{figure*}[!t]
\centering
\includegraphics[width=6.5in]{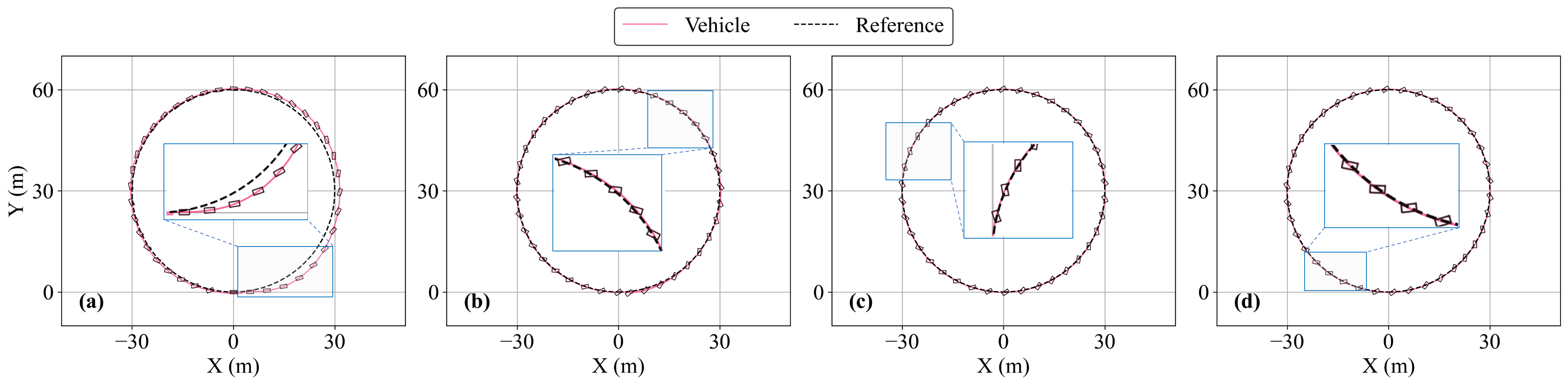}
\caption{Comparison of vehicle paths and reference path with a 30-meter radius circle. (a) 0-20 sec. (b) 20-40 sec. (c) 40-60 sec. (d) 60-80 sec. }
\label{circle30}
\end{figure*}

\begin{figure}[!t]
\centering
\includegraphics[width=3.3in]{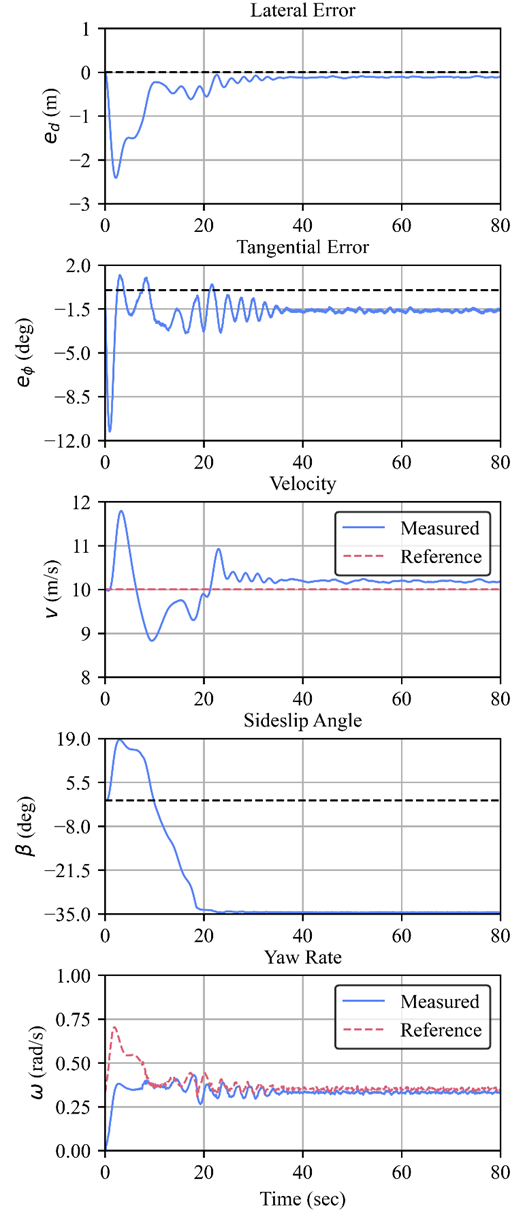}
\caption{Comparison of vehicle paths and reference path with a 30-meter radius circle.}
\label{circle30results}
\end{figure}

\begin{figure}[!t]
\centering
\includegraphics[width=3.3in]{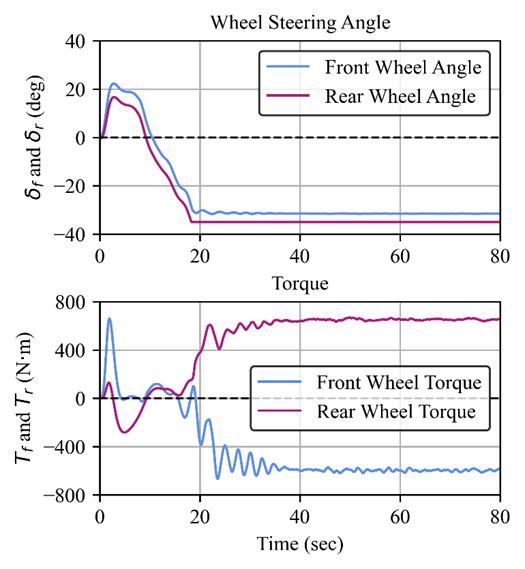}
\caption{Control inputs of the drifting controller.}
\label{circle30input}
\end{figure}

The proposed drifting controller demonstrates effective tracking of the reference trajectory, as illustrated in Fig. \ref{circle30} and Fig. \ref{circle30results}, with the corresponding wheel steering angle and torque control commands depicted in Fig. \ref{circle30input}. The maximum lateral error is -2.41 m, the root-mean-square (RMS) lateral error is -0.31 m, and the steady-state lateral error is -0.11 m. Considering that the controller without error compensation exhibits a steady-state lateral error of nearly -1.5 m, this performance is deemed acceptable.

At 10 seconds into the simulation, the vehicle’s sideslip angle shifts from positive to negative, while the yaw rate remains positive (positive for counterclockwise and negative for clockwise). This marks the onset of a drifting state, where the vehicle transitions into controlled lateral motion. From 20 seconds to 80 seconds, the vehicle remains in a stable drifting state, with a sideslip angle of 35 degrees and a yaw rate of 0.33 rad/s.

It can be seen from Fig. \ref{circle30input} that the front and rear steering angles exhibit similar behavior during drifting, while the tor-ques act in opposite directions. Together, the steering angles and torque distribution generate an extra counterclockwise yaw moment, helping the 4WD-4WS vehicle maintain the drift and stabilize its trajectory.

As illustrated in the preceding figures, the transition of a 4WD-4WS high-speed vehicle from typical driving to drifting state is categorized into three distinct phases. Initially, concurrent steering of both front and rear wheels is per-formed with the simultaneous application of driving torque.  Subsequently, in the second phase, the initiation of counter-steering in both front and rear wheels occurs, accompanied by an increase in rear driving torque and a corresponding de-crease in front driving torque.  Finally, steady-state drift is attained in the third phase by maintaining a slightly smaller front wheel angle compared to the rear wheel angle.

\begin{figure}[!t]
\centering
\includegraphics[width=3.3in]{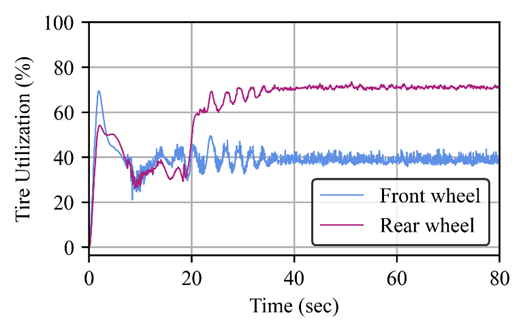}
\caption{Tire utilization.}
\label{circle30tireuse}
\end{figure}

Fig. \ref{circle30tireuse} shows the variation in tire utilization for the front and rear wheels. Tire force utilization refers to the ratio of the actual tire force to the tire's maximum possible force, providing a measure of how effectively the tires contribute to maintaining vehicle dynamics. Compared to RWD vehicles, where the rear wheels are almost always saturated during drifting, the rear tire utilization of the 4WD-4WS vehicle remains below 80\%. This suggests that the 4WD-4WS system does not rely on inducing rear-wheel slip to achieve drifting, unlike traditional RWD vehicles. Instead, the system distributes forces more evenly across all four tires, which allows the vehicle to maintain greater stability and control without over-loading the rear tires.

Furthermore, this more balanced distribution of forces means that the 4WD-4WS vehicle experiences lower tire wear during drifting compared to other vehicle types. Since the rear tires are not excessively stressed, their longevity is significantly improved, resulting in reduced tire wear and maintenance costs. This makes the 4WD-4WS system not only more efficient in maintaining vehicle dynamics but also gentler on the tires, ensuring better performance over time.

\subsection{Variable Curvature Circular}

Following the analysis of constant curvature trajectories, this section examines the variable curvature circular trajectory, which introduces more complexity and better represents real-world conditions in dynamic motion. 

\begin{figure}[!t]
\centering
\includegraphics[width=3.3in]{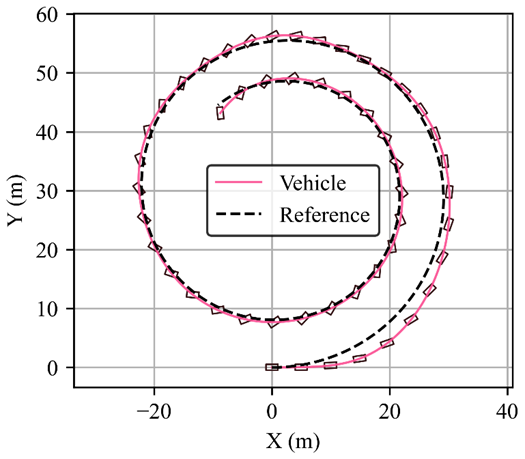}
\caption{Comparison of vehicle paths and the reference path with a variable curvature circular trajectory.}
\label{spiralresults}
\end{figure}

The curvature of the circular trajectory ranges from 1/30 to 1/18 over a distance of 240 m. From Fig. \ref{spiralresults}, it can be ob-served that the vehicle closely tracks the reference trajectory, with only minor deviations resulting from the changing curvature. Moreover, the vehicle is able to maintain a sustained drift, highlighting the robustness of the control system.

\section{CONCLUSION}
This article proposes a hierarchical drift controller for 4WD-4WS vehicles that simplifies drift control by eliminating the need for precomputed drift equilibrium.
First, a feedback model for yaw rate and trajectory tracking is designed, enabling the vehicle to maintain control under large sideslip angles and enhancing its handling limits. Next, the upper layer uses an MPC framework that accounts for disturbances from inverse tire model errors. Then, the lower layer employs a Newton-Raphson-based method to solve the inverse Magic Formula tire model, determining steering and torque inputs. Finally, circular trajectory simulations confirm the controller's effectiveness, demonstrating not only precise and stable drift control but also highlighting the advantages of 4WD-4WS vehicles in drifting compared to other vehicle configurations. This work lays the foundation for advancing 4WD-4WS vehicle dynamics. However, this study does not explore the controller’s performance on other general paths, leaving room for future research.
Future research will focus on two key areas. The first is achieving efficient transitions between conventional driving and drifting, ensuring seamless control across different driving modes. The second is developing trajectory planning methods specifically tailored for drift control, aiming to minimize lap times and enhance performance in racing scenarios. These directions will further unlock the potential of 4WD-4WS vehicles in dynamic and high-performance applications.

\bibliographystyle{unsrt}  

\bibliography{references}

\end{document}